\documentclass[a4paper,11pt]{article}
\pdfoutput=1
\usepackage{jcappub}
\title{Extending BEAMS to incorporate correlated systematic uncertainties}

\author[a,b]{Michelle Knights,} 
\author[a,b,c]{Bruce A. Bassett,} 
\author[a,d]{Melvin Varughese,} 
\author[e]{Ren\'ee Hlozek,} 
\author[f]{Martin Kunz,} 
\author[g]{Mat Smith} 
\author[a,b]{and James Newling}

\affiliation[a]{African Institute for Mathematical Sciences, 6 Melrose Road, Muizenberg, 7945, South Africa}
\affiliation[b]{Department of Mathematics and Applied Mathematics, University of Cape Town, Rondebosch, Cape Town, 7700, South Africa}
\affiliation[c]{South African Astronomical Observatory, Observatory Road, Observatory, Cape Town, 7935, South Africa}
\affiliation[d]{Department of Statistical Sciences, University of Cape Town, Rondebosch, Cape Town, 7700, South Africa}
\affiliation[e]{Department of Astrophysical Sciences, Princeton University, Princeton, NJ 08544, USA}
\affiliation[f]{D\'epartement de Physique Th\'eorique and Center for Astroparticle Physics, Universit\'e de Gen\`eve, Quai E.\ Ansermet 24, CH-1211 Gen\`eve 4, Switzerland} 
\affiliation[g]{Department of Physics, University of the Western Cape, Bellville,Cape Town, 7535, South Africa}

\emailAdd{michelle.knights@gmail.com}

\date{}

\abstract{New supernova surveys such as the Dark Energy Survey, Pan-STARRS and the LSST will produce an unprecedented number of photometric supernova candidates, most with no spectroscopic data. Avoiding biases in cosmological parameters due to the resulting inevitable contamination from non-Ia supernovae can be achieved with the BEAMS formalism, allowing for fully photometric supernova cosmology studies. Here we extend BEAMS to deal with the case in which the supernovae are correlated by systematic uncertainties. The analytical form of the full BEAMS posterior requires evaluating $2^N$ terms, where $N$ is the number of supernova candidates. This `exponential catastrophe' is computationally unfeasible even for $N$ of order 100. We circumvent the exponential catastrophe by marginalising numerically instead of analytically over the possible supernova types: we augment the cosmological parameters with nuisance parameters describing the covariance matrix and the types of all the supernovae, $\tau_i$, 
that we include in our MCMC analysis. We show that this method deals well even with large, unknown systematic uncertainties without a major increase in computational time, whereas ignoring the correlations can lead to significant biases and incorrect credible contours. We then compare the numerical marginalisation technique with a perturbative expansion of the posterior based on the insight that future surveys will have exquisite light curves and hence the probability that a given candidate is a Type Ia will be close to unity or zero, for most objects. Although this perturbative approach changes computation of the posterior from a $2^N$ problem into an $N^2$ or $N^3$ one, we show that it leads to biases in general through a small number of misclassifications, implying that numerical marginalisation is superior.}


\begin{document}
\maketitle

\renewcommand{\floatpagefraction}{0.7}

\section{Introduction}
Type Ia supernovae (SN Ia) are standardisable candles, making them one of the most reliable distance measures and a cornerstone of cosmology ever since the discovery of the late time accelerated expansion of the Universe \cite{riess,perlmutter}.
\\
 Future surveys which will produce large amounts of photometric data, such as the Dark Energy Survey (DES) \cite{des}, Pan-STARRS \cite{pan} and the Large Synoptic Survey Telescope (LSST) \cite{lsst}, will increase the number of SN Ia candidates by orders of magnitude. While a foolproof method of identifying a Type Ia is to analyse its observed spectrum, taking spectra is expensive and, for surveys such as those mentioned above, it will be unfeasible to perform spectroscopic follow-up for all candidates, introducing a possible bias due to contamination from Type Ib/c and Type II supernovae, which we collectively denote non-Ia supernovae (SN nIa) \cite{hlozek,falck}. 
 However, using the photometric information gathered by the survey, one can fit template light curve models to the data using a template fitter such as MLCS2k2 \cite{mlcs}, SALT2 \cite{salt} or a model fitter such as in \cite{sako}, which gives a probability for each object to be a Type Ia. 
\\
In previous surveys, the relative SN Ia/nIa probabilities were used only to determine candidates for spectroscopic follow-up \cite{sdss2}. Without spectroscopic follow-up, applying a probability cut (for example, taking all supernovae with probability greater than 0.9 to be a Type Ia) will introduce a bias in the cosmological parameters \cite{hlozek}. To avoid such biases, one can either demand a very high purity, which excludes much of the data \cite{gjergo, bernstein,campbell} or use all the data within a statistical framework that accounts for the contamination. One such  method, developed by Kunz, Bassett \& Hlozek \cite{kunz} is Bayesian Estimation Applied to Multiple Species (BEAMS). BEAMS has recently been applied to the full three years of data from the SDSS-II supernova survey \cite{sdss1,sdss2}, which reduced the $\Omega_m-\Omega_\Lambda$ contours by a factor of three relative to the spectroscopic data alone \cite{hlozek}.
\\
Despite this success, the current implementation of BEAMS assumes the supernovae are not correlated with each other; an approach which will not be appropriate for
future surveys. In general, systematic uncertainties are correlated and, as the number of discovered supernovae increases, we are entering an era where systematic uncertainties are comparable to statistical uncertainties. This means the off-diagonal terms of the covariance matrix cannot be ignored. To analytically account for correlations between supernovae in the BEAMS posterior requires summing over $m^N$ terms, where $N$ is the number of supernova candidates and $m$ is the number of possible supernova types, which here we take to be two, corresponding to Ia's and nIa's. Clearly, this is computationally impossible, but in this paper we will show that if the form of the covariance matrix is known (for example, if the known sources of correlations can be parameterised), BEAMS can still be used to estimate cosmological parameters in an unbiased way, using a numerical marginalisation over supernova type.
\\
After a brief review of the theory of BEAMS in section \ref{beams_sec}, we move on to discuss possible sources of correlations in section \ref{correlations_sec}. We introduce mock supernova datasets described by three different covariance matrices in section \ref{cov_mat_sec}. In section \ref{n_beams} we discuss a solution to the `exponential catastrophe' of the correlated form of the BEAMS posterior using numerical marginalisation of types and in section \ref{p_beams} we compare it with an alternative solution based on a perturbative expansion of the BEAMS posterior.

\section{BEAMS}
\label{beams_sec}
Cosmological parameter estimation usually proceeds by maximising the posterior, $P(\theta|D)$, where $D$ is the set of redshifts and distance moduli of spectroscopically confirmed Type Ia supernovae and $\theta$ is the set of cosmological parameters, such as $\Omega_m$, $\Omega_\Lambda$ and $H_0$. What happens if we do not have spectroscopic confirmation of an object's type but only a probability that it is a Ia? Unbiased parameter estimation in this case can be achieved using Bayesian Estimation Applied to Multiple Species (BEAMS).
\\
BEAMS \cite{kunz} considers all data in a given sample, appropriately weighting each data point based on its probability of being a Ia. Let $\tau_i$ be the type of object $i$ and $\tau_i=\rm{Ia}$ if the object is a Type Ia and $\tau_i=\rm{nIa}$ if the object is not a Ia (for example, if it is a Type Ib/c or a Type II supernova). Then we can write the posterior, $P(\theta|D)$ (here $\theta$, $D$ and $\tau$ are understood to be vectors), as
\begin{equation}
 P(\theta|D) = \displaystyle\sum\limits_{\tau} P(\theta,\tau|D).
\end{equation}
This sum marginalises over all possible combinations of types for the dataset so $\tau$ here is a length-$N$ vector (where $N$ is the number of objects). For example, if there were three objects in the dataset, the first term in the sum would have $\tau_1=\rm{Ia}$, $\tau_2=\rm{Ia}$, $\tau_3=\rm{Ia}$, the second term would have $\tau_1=\rm{nIa}$, $\tau_2=\rm{Ia}$, $\tau_3=\rm{Ia}$ etc. Thus, in general, for the case of two distinct object types, this is a $2^N$ summation. Applying Bayes' theorem gives that
\begin{equation}
 P(\theta,\tau|D) = P(D|\theta,\tau) \frac{P(\theta,\tau)}{P(D)}.
\end{equation}
$P(D)$, the Bayesian evidence, can be considered as a normalisation factor and ignored in further calculations. We will assume that $P(\theta,\tau) = P(\theta)P(\tau)$ (see \cite{kunz} for a discussion of this assumption). $P(\theta)$ is the usual prior on the parameters (probability based on prior knowledge about the parameters), and $P(\tau)$ can be written as
\begin{equation}
 P(\tau) = \displaystyle\prod\limits_{\tau_j=\rm{Ia}}P_j \displaystyle\prod\limits_{\tau_k=\rm{nIa}}(1-P_k).
\end{equation}
This is the product of the probabilities, $P_i$, for all the objects typed as a Ia, $\tau_i=\rm{Ia}$, multiplied by the product of $(1-P_i)$ for all the objects typed as nIa's, $\tau_i=\rm{nIa}$. Here we assume the data and the object types are uncorrelated (see \cite{newling} for details of BEAMS without this assumption). Thus the BEAMS posterior is given by
\begin{equation}
\label{beams_eq}
 P(\theta|D) \propto P(\theta) \displaystyle\sum\limits_{\tau} P(D|\theta,\tau) \displaystyle\prod\limits_{\tau_j=\rm{Ia}} P_j \displaystyle\prod\limits_{\tau_k=\rm{nIa}}(1-P_k).
\end{equation}
\\
As an order $2^N$ calculation, this is computationally unfeasible. In the case of uncorrelated data, there is a simplification we can use to make the problem tractable. By decomposing the likelihood as a product of probabilities, it can be shown \cite{kunz} that the posterior can be written as
\begin{equation}
\label{beams}
 P(\theta|D) \propto P(\theta) \displaystyle\prod\limits_{i=1}^N \bigg(\mathcal{L}_{\rm{Ia},i}(\theta) P_i + \mathcal{L}_{\rm{nIa},i}(1-P_i)  \bigg),
\end{equation}
where $\mathcal{L}_{\rm{Ia},i}$ is the likelihood assuming the $i$'th object is a Ia and $\mathcal{L}_{\rm{nIa},i}$ is the likelihood assuming it is a nIa.
\\
This formula works well for uncorrelated data but is unlikely to be accurate in the case of correlated data. For the large datasets which will be available in the future, correlations among the data cannot be ignored and another solution must be found to apply BEAMS to these datasets in a computationally feasible way.

\section{Correlated supernova data}
\label{correlations_sec}
Correlations between supernovae has only recently become an issue that must be included in cosmological parameter sets (e.g. \cite{conley,amanullah}) but will become progressively more important as we push on the systematics floor related to SNIa surveys.
Correlated systematic uncertainties, the focus of this paper, can arise from a large number of sources and a detailed study of correlations has yet to be undertaken due to the diversity and complexity of the various contributing factors. Schematically correlations can arise from:
\begin{itemize}
\item {\em Peculiar velocities:} When supernovae are within 50 Mpc or so of each other the peculiar velocities of their host galaxies will be correlated by  large-scale bulk flows. These peculiar velocities cause correlated redshift errors. Usually redshift errors are converted into additional distance modulus errors (see e.g. \cite{kessler}) but even if this is not done it will cause errors in the recovered cosmological parameters. See, for example, \cite{hui,cooray,gordon,davis}.

\item {\em Redshift-colour and redshift-stretch correlations:} If there is no spectroscopic host galaxy redshift for an object, the
redshift is estimated photometrically either from the host or the supernova multi-band light curves.
The effect of redshift on the light curve is degenerate to some extent with the stretch
and colour corrections. Hence errors on redshift will correlate with those in 
the colour and stretch and thus with the estimated distance modulus of the supernova \cite{snls}.

\item {\em Filter errors:} Transmission curves for the actual filters used on a telescope approximate true through-put. Errors in measuring these transmission curves (or time dependent changes of the filters)\cite{doi} will tend to induce redshift-dependent correlations. For example, in \cite{feindt}, the authors consider the error in the flux calibration of the telescope which, since it affects each filter differently, correlates objects at similar redshifts. Zero-point photometry errors are a major source of uncertainty in current surveys that are common to all supernovae observed with the same telescope; e.g. \cite{amanullah}.
 
\item {\em Template error correlations:} Unaccounted for evolution of supernovae with redshift causes correlated errors due to errors in the light curves that form the training set. An example of this is the `U-band anomaly' which causes discrepancies between the SALT2 and MLCS2k2 light curve fitters which may be related to an excess of flux in the UV at high redshift in SNIa \cite{foley, kessler}. 

\item {\em Observational conditions:} Bad weather will cause holes in the light curve coverage of all supernovae visible at a given time, while seeing conditions will alter photometry measurements in a correlated way. These will induce subtle correlations between objects observed on the same night in similar conditions \cite{mandel,lsst}. 

\item {\em Combining data from multiple telescopes:} The covariance matrix for combining data from multiple telescopes can be very complex, as discussed in the 3 year SNLS analysis \cite{conley}. With the exception of perhaps the final LSST dataset, combining data from multiple surveys will continue to be standard. 

\item {\em Gravitational lensing:} Supernovae that are close together on the sky, at similar redshifts will experience similar brightening or dimming due to lensing, depending on the matter distribution along the line of sight. Future large surveys will have mass maps and hence will be able to predict and remove this signal to some extent \cite{gunnarsson, jonsson, amendola}.

\item {\em Dust:} Supernovae along neighbouring lines of sight will suffer similar extinction from the Milky Way and from any intergalactic dust, which will induce correlations \cite{zhang,corasaniti}.

\item {\em Host-galaxy correlations:} There is now solid evidence that dispersion in the Hubble diagram correlates with the properties of host galaxies, particularly the galaxy type, size and mass. See, for example \cite{kelly,gupta,dandrea,meyers}.

\item {\em Spectroscopic targeting correlations:} Since spectroscopic follow-up is typically not random, there may be hidden correlations. For example, follow-up may favour candidates well-separated from the host galaxy core. Malmquist bias can also cause correlations which depend on the details of the spectroscopic survey \cite{sdss2,kessler}. If there is an unknown systematic that such objects are intrinsically brighter/fainter than average, this will cause a correlated systematic error \cite{sdss2}.

\item {\em nIa correlations:} Many of the sources of correlation listed here will also affect nIa's, causing correlations between their distance moduli. However, these correlations will typically be much smaller than the intrinsic dispersion of the nIa population.

\end{itemize}
\noindent
Given the complexity of these various effects, developing a `realistic' correlation matrix is beyond the scope of this work and will need to be laboriously built from simulations and detailed studies. Here we wish instead to develop ways of dealing with the general problem posed by the $2^N$ `exponential catastrophe' that correlations present. We use several toy model covariance matrices to study potential resolutions of this catastrophe, finding that numerical marginalisation over the supernova types performs best. 

\section{Mock data}
\label{cov_mat_sec}
In order to determine the effect that correlated data can have on parameter estimation with BEAMS, we use mock supernova datasets, the properties of which are known. This enables us to estimate the magnitude of the bias introduced by correlations between the data points and to determine the optimum way to handle this additional source of error. 
\\
\noindent
To simulate supernova data, we need to create a distance modulus $\mu(z)$ for each object. The distance modulus is defined as 
\begin{equation}
 \mu(z) = m - M = 5 \text{log}_{10}\left(\frac{d_L}{1Mpc}\right) + 25,
\end{equation}
where $m$ is the apparent magnitude of the object, $M$ is the absolute magnitude of the object and $d_L$ is the luminosity distance to the object in Mpc \cite{dodelson}. Through the distance modulus, SN Ia can be used to constrain cosmological parameters. The luminosity distance depends on the cosmological parameters (assuming a $\Lambda$CDM model) by
\begin{equation}
 d_L (z) = \frac{c(1+z)}{H_0 \sqrt{-\Omega_k}} \text{sin}\left( H_0 \sqrt{-\Omega_k} \int \frac{dz'}{H(z')} \right),
\end{equation}
where 
\begin{equation}
 H(z) = H_0 \bigg(\Omega_m (1+z)^3 + \Omega_\Lambda + \Omega_k(1+z)^2 \bigg)^{1/2},
\end{equation}
and $H_0$ is the Hubble constant, $\Omega_m$ is the energy density of matter, $\Omega_\Lambda$ is the energy density of dark energy and $\Omega_k$ is the energy density of curvature (all densities relative to the critical density) \cite{dodelson}. We used a $\Lambda$CDM model to generate the mock data with the following parameters: $H_0=70.4$ km/s/Mpc, $\Omega_m=0.272$ and $\Omega_\Lambda=0.628$ (values taken from \cite{komatsu}).
\\

\begin{figure}[th]
\vspace{-30pt}
 \centering
\includegraphics[width=0.65\columnwidth]{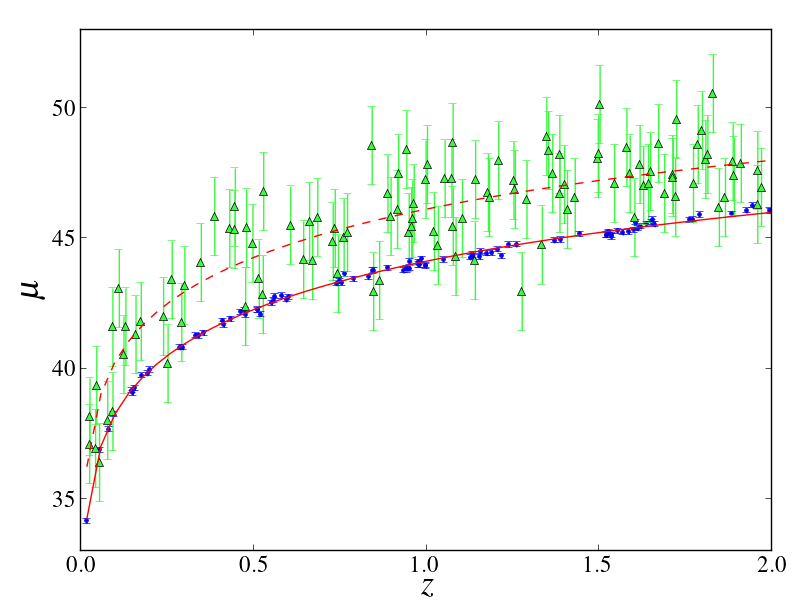}
\caption{Distance modulus ($\mu$) as a function of redshift for a typical mock dataset with 200 supernovae. Blue points are SN Ia, green triangles are SN nIa and the red line is the fiducial cosmological model. The dotted line is the model for the nIa's. The intrinsic dispersion of the Ia's is $\sigma_{\text{Ia}}=0.1$ and that of the nIa's is $\sigma_{\text{nIa}}=1.5$, as indicated by the error bars. We show the true types and intrinsic scatter of the points for clarity here but note that neither are given to BEAMS, which receives only the probabilities and the $\mu$ values as input.}
\label{data}
\end{figure}
\vspace{-7pt}
\noindent

\begin{figure}[ht!]
 \centering
 \includegraphics[width=0.6\columnwidth]{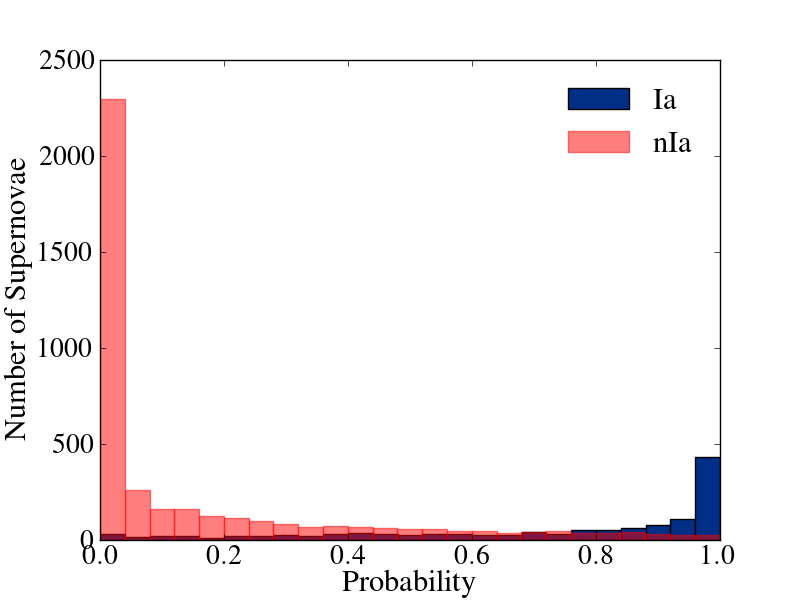}
 \caption{Histogram of Dark Energy Survey-like SNIa probabilities, where the supernova data was generated using SNANA and fitted with MLCS2k2 to obtain the probabilities. The blue (dark), outlined bars are Ia's and the red (light) bars are nIa's. This figure illustrates the point that future data will deliver bimodal probabilities peaked near $0$ and $1$, and helps guide the probability distribution for our mock data.}
 \label{des}
\end{figure}

SN Ia typically have very little scatter in their distribution of distance moduli, while nIa's tend to be widely scattered. SN Ia also tend to be brighter than nIa's. As such, we model the mock data as drawn from two populations: the Ia distribution is a narrow Gaussian centred on the fiducial cosmological distance modulus, $\mu(z)$, with standard deviation $\sigma_{\text{Ia}}$. The nIa distribution is a wide Gaussian, centred on $\mu(z)+b$, where $b$ is a constant shift\footnote{It was shown in \cite{hlozek} that BEAMS is able to reconstruct the correct redshift evolution for the nIa distribution. Here we assume, without loss of generality, that $b$ is a constant shift.}. The nIa distribution has standard deviation $\sigma_{\text{nIa}}$. While in general one could model each known type of supernova as a separate population, two populations are sufficient for this work. For each object, redshifts in the range of $0.01<z<2.0$ were drawn from a uniform random distribution, while probabilities, $P_i$, were drawn from a 
distribution that simulates the expected Dark Energy Survey probability distribution, shown in figure \ref{des}. There were roughly equal numbers of Ia's and nIa's. Here we assume that the probabilities are correct, but it was shown in \cite{kunz,hlozek} that BEAMS can deal with biased probabilities in general.
\\
\\The type of each object was randomly chosen with probability $P_i$ which allowed the object to be assigned a distance modulus. Residuals were drawn from the appropriate distribution, depending on type, correlated using one of the covariance matrices described below and added to the appropriate Ia or nIa mean. A typical mock dataset distance modulus diagram is shown in figure \ref{data}, where we have chosen the standard deviation of the Ia population to be $\sigma_{\text{Ia}}=0.1$, that of the nIa's to be $\sigma_{\text{nIa}}=1.5$ and the shift of the nIa population to be $b=2.0$. In this paper we experimented with datasets varying between 200 and 1000 supernovae.
\\
Since the exact form of the true covariance matrix for supernovae is unknown, we use simple toy models for a covariance matrix for the mock data. Our main goal is to show how BEAMS can be used for correlated data, so we do not attempt to model a realistic covariance matrix. We analyse the effects of three different covariance matrices. In all cases, we order the objects according to redshift, since we expect the strongest correlations to be functions of redshift. 
\\
As they will come from the same sources, we expect Ia-nIa and nIa-nIa correlations to be of similar magnitude as Ia-Ia correlations. However, since the dispersion of the nIa distribution is so large, these small correlations have little influence and only the Ia-Ia correlations affect the results. We found that when we created a dataset with correlations between all supernova types and analysed it assuming only Ia correlations, the results were identical to an analysis which included non-zero nIa-nIa and Ia-nIa correlations. Hence, we only consider Ia-Ia correlations in our analysis.

\subsection{Wedding cake covariance matrix}
This example matrix is based on the supernova covariance matrix in Kim et al. \cite{kim}, which is guaranteed to be positive-definite. This matrix has a wedding cake or step-like structure: an error entered at a given redshift contributes to the error of all higher redshift objects. This structure naturally arises as one loses observational features at higher redshifts thus introducing errors. We constructed a matrix based on this, in which we divided the data set into five redshift bins, adding an extra source of error in each bin. This structure is given as

\begin{equation}
 C_{ij} = \sigma_i \sigma_j \delta_{ij} + V_{ij},
 \label{wedding}
\end{equation}

\noindent
where $\sigma_i=\sigma_{\text{Ia}}$ if $\tau_i$=Ia and $\sigma_i=\sigma_{\text{nIa}}$ if $\tau_i$=nIa and

\begin{equation}
 V_{ij} = 
\begin{cases}
 \displaystyle \sum \limits_{k=1}^{n_{i,j}} s_k & \text{if $\tau_i$=$\tau_j$ = Ia} \\
0 & \text{otherwise},
\end{cases}
\end{equation}

\noindent
where $n_{i,j}$ is the bin to which the object belongs. To produce the step-like structure, $n_{i,j}=\lfloor \frac{min(i,j)}{N/5} \rfloor +1$ (where ``$\lfloor \rfloor$'' indicates the floor function, rounding down to the nearest integer). For this covariance matrix we used $s_k=0.015$ for $k=1$ to $5$, as illustrated in the top left panel of figure \ref{correlated}.

\subsection{Decaying Covariance Matrix}
This covariance matrix assumes positive correlations between objects which are nearby in the covariance matrix, with the correlations decaying as the distance between the indices of objects is increased. The exact form of this covariance matrix is
\begin{equation}
 C_{ij} = 
\begin{cases}
 \sigma_{\text{Ia}}^2 & \text{if i=j $\&$ $\tau_i$=$\tau_j$ = Ia} \\
 \sigma_{\text{nIa}}^2 & \text{if i=j $\&$ $\tau_i$=$\tau_j$ = nIa} \\
 \frac{x \sigma_{\text{Ia}}^2}{|i-j|} & \text{if i$\neq$j $\&$ $\tau_i$=$\tau_j$ = Ia} \\
 0 & \text{otherwise} \\
 
\end{cases}
\end{equation}
This is illustrated in the middle left panel of figure \ref{correlated}, where we set $x=0.7$.

\subsection{Block-diagonal covariance matrix}
To illustrate our method with a more realistic example, we used the covariance matrix for the Union2 sample of 557 supernovae. \cite{amanullah,suzuki} constructed this matrix by parameterising all known sources of correlations for the Union2 dataset, fitting these nuisance parameters simultaneously with the cosmological parameters and providing an estimate of the best-fit covariance matrix. We binned into 11 redshift bins, which we then applied to our mock data. An example for one mock data realisation can be seen in the bottom left panel of figure \ref{correlated}. We have set nIa-nIa and nIa-Ia correlations to zero, since these should be very small compared to the intrinsic dispersion of the nIa's. We added a $\sigma_i^2$ term to the diagonal where $\sigma_i=\sigma_{\text{Ia}}$ if object $i$ is a Ia and $\sigma_i=\sigma_{\text{nIa}}$ if it is a nIa.
The block diagonal structure arises largely from the fact that supernovae from the same survey are correlated with each other. 

\newpage
\section{Numerical marginalisation over supernova type}
\label{n_beams}
\subsection{Theory}
A solution to the computational problem of handling correlated data would be to perform the marginalisation over the types numerically instead of analytically, allowing us to use the correlated BEAMS posterior without the $2^N$ sum. In order to do this, we create $N$ discrete nuisance parameters, the types of the supernovae, $\tau_i$, and marginalise over these parameters in our Markov Chain Monte Carlo (MCMC) \cite{metropolis, hastings} analysis. This problem is similar to the Ising spin problem because these parameters are discrete and can only assume one of two values \cite{ising}.
\\
In each step of the MCMC chain, we randomly select one object and set the corresponding $\tau_i=$Ia with probability $P_i$, which significantly speeds up convergence when compared with varying all the types each step.  To further improve convergence, we choose the initial type parameters based on the ratio of Ia to nIa uncorrelated likelihoods for each object. If the ratio is greater than one, we set the initial type for that object to a Ia, otherwise to a nIa. To compute this initial likelihood ratio, we use some fiducial values for the cosmological parameters. This initial choice for the types has little bearing on the final result, but without it, the chain typically starts with a very low likelihood because many objects have the wrong type and it takes much longer to reach the region of the peak. Thus, if $\theta$ represents the set of parameters in the MCMC chain and $\tau$ the types, the usual MCMC acceptance criterion is
\begin{equation}
 R = \text{min} \left(\frac{\mathcal{L}(\theta_{\text{new}},\tau_{\text{new}})}{\mathcal{L}(\theta,\tau)}, 1 \right)
\end{equation}

We applied this technique using MCMC to estimate cosmological parameters from the mock datasets. We ran at least three MCMC  chains for each dataset produced and ensured all chains were converged using the Gelman-Rubin \cite{gelman} criterion for convergence, with $R<1.01$. Chains were, on average, 60 000 steps long after burn-in was removed. It should be noted that for all MCMC chains, the following flat, wide priors were used, to ensure unbiased parameter estimation (since small datasets were used to reduce computational time, which tend to have large contour areas): $-0.2<\Omega_m<1.2$, $-0.2<\Omega_\Lambda<1.2$ and $10<H_0< 130$. We allowed the shift, $b$, of the nIa population to vary with a flat prior of $1<b< 3$ and the standard deviations of the populations, $\sigma_{\text{Ia}}$ and $\sigma_{\text{nIa}}$ to vary in log space with priors of $-3.0<\text{log}(\sigma_{\text{Ia}})<-1.2$ and $-0.7<\text{log}(\sigma_{\text{nIa}})<1.5$. To allow for the most general case where only the form of the covariance matrix is known, but its 
parameters are not, we also varied the covariance parameters for each individual covariance matrix over a flat prior, except for the block covariance matrix whose off-diagonal terms remained fixed, since they were determined from the Union2 dataset.

\subsection{Testing BEAMS with numerical marginalisation}
We first generated uncorrelated datasets to directly compare the numerical and analytic marginalisation of BEAMS. We found that the $\Omega_m-\Omega_\Lambda$ credible contours\footnote{In this paper, we use credible intervals, the Bayesian analogue to confidence intervals. Bayesian credible intervals represent the degree of belief in the parameter estimates, and are derived from the posterior probability. They are not, in general, equivalent to frequentist confidence intervals \cite{jaynes}.} produced using eq.(\ref{beams}) and those produced using the numerical marginalisation technique were identical for the uncorrelated datasets, showing that the technique works as expected.
\\

\begin{figure}[phtb!]
\vspace{-30pt}

\begin{minipage}{0.5\textwidth}
\centering
 \includegraphics[width=0.99\columnwidth]{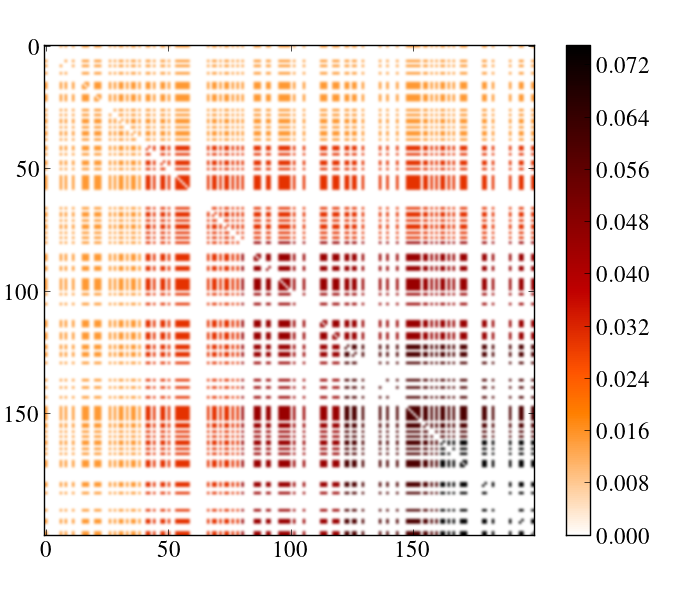}

\end{minipage}
\hspace{0.2cm}
\begin{minipage}{0.5\textwidth}
\centering

\includegraphics[width=0.99\columnwidth]{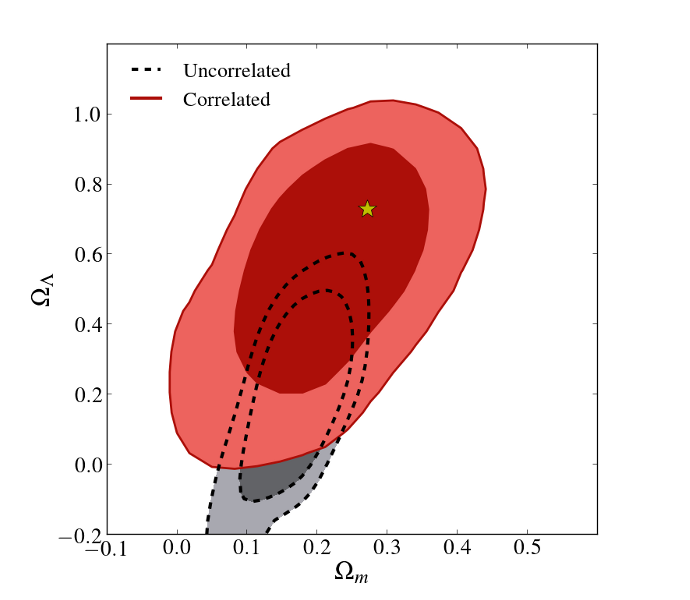}
 \end{minipage}

\vspace{10pt}
\begin{minipage}{0.5\textwidth}
\centering
 \includegraphics[width=0.99\textwidth]{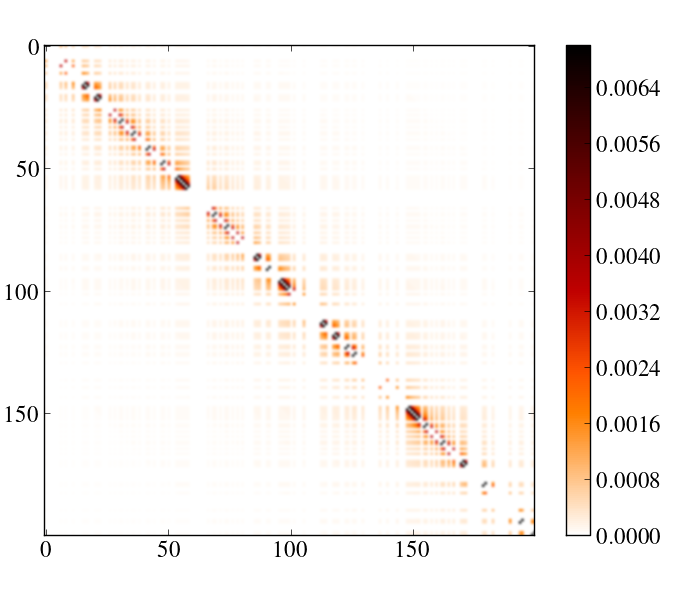}
 \end{minipage}
\hspace{0.2cm}
\begin{minipage}{0.5\textwidth}
\centering

\includegraphics[width=0.99\columnwidth]{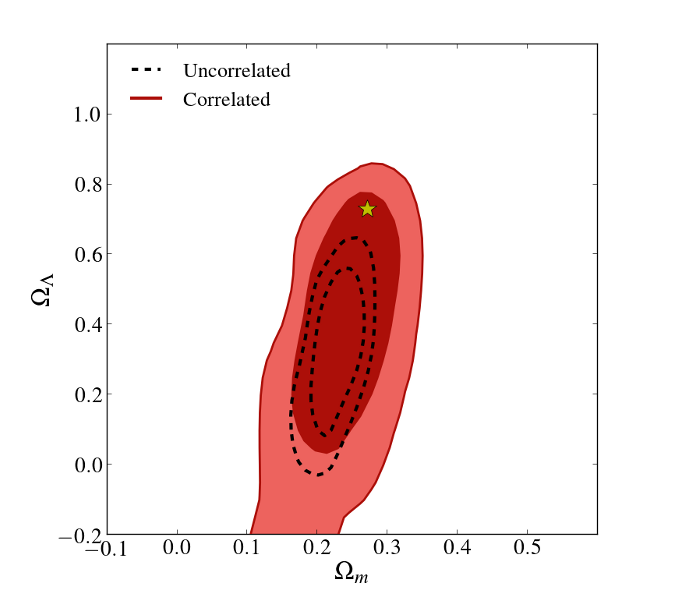} 
\end{minipage}

 \begin{minipage}{0.5\textwidth}
\vspace{10pt}
\centering
 \includegraphics[width=0.99\textwidth]{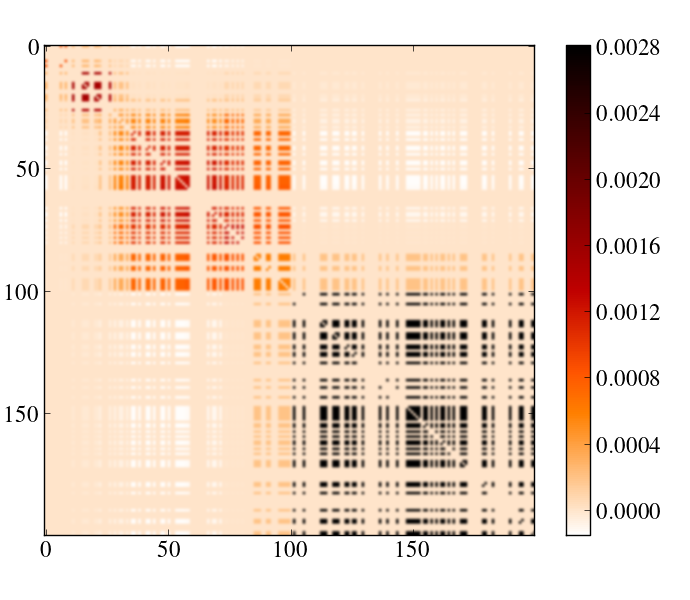}
 \end{minipage}
\hspace{0.2cm}
\begin{minipage}{0.5\textwidth}
\centering

\includegraphics[width=0.99\columnwidth]{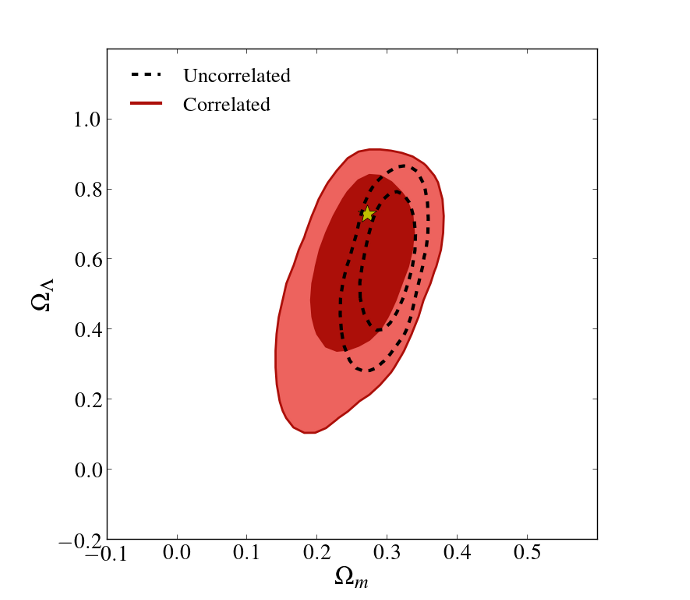}
\end{minipage}

\caption{\emph{Results of the numerical marginalisation of types with BEAMS using three different covariance matrices:} The left column shows schematic views of the three covariance matrices, where the colour indicates the amount of correlation between two data points. The right column shows the $\Omega_m - \Omega_\Lambda$ contours, marginalising over $H_0$, when the correlated data sets are analysed assuming no correlations (dotted lines) and using correlated BEAMS (filled contours). The covariance matrices and their parameters are (from top to bottom): the wedding cake covariance matrix ($s_k$=0.015 for $k$=1 to 5), the decaying covariance matrix ($x$=0.7) and the block diagonal covariance matrix. All covariance parameters (for the wedding cake and decaying matrices) are marginalised over with wide, flat priors. Note: for all matrices, the diagonal has been removed for clarity because $\sigma_{\text{nIa}}$ is so much larger than the off-diagonal terms.}

\label{correlated}
\end{figure}

\noindent
It is interesting to note that the marginalisation method is faster than uncorrelated BEAMS, since there are half the calculations to perform at each step (uncorrelated BEAMS computes the likelihood assuming the object is nIa and Ia respectively, whereas the numerical marginalisation only requires the likelihood given the current types). However, it takes longer for the correlated BEAMS chains to converge. On average, using the Gelman-Rubin \cite{gelman} criterion for convergence, uncorrelated BEAMS chains converge within 10 000 steps whereas the correlated BEAMS chains only converge after about 40 000 steps. We tested datasets of up to 1000 data points in size and found that, although the amount of time taken to do a single likelihood calculation increases linearly, the \emph{number of steps} taken to converge does not increase appreciably as the number of data points is increased. Hence, computational complexity will not be an issue for correlated BEAMS.
\\
Figure \ref{correlated} compares the uncorrelated BEAMS approach with correlated BEAMS (using numerical marginalisation of types) in the case of correlated data, based on the three different covariance matrices from section \ref{cov_mat_sec}. The uncorrelated approach includes no information about correlations and hence it is not surprising that it is biased at $>2\sigma$ for both the decaying covariance matrix and the wedding cake covariance matrix. In contrast, correlated BEAMS with numerical marginalisation correctly estimates the cosmological parameters without any bias. There is less of an effect from the block diagonal covariance matrix, because both the correlations and the dataset are small. Although even in this case, it is clear the uncorrelated BEAMS contours appear to be mildly biased.

\cleardoublepage
\subsection{Testing the Contours and Coverage Properties of the BEAMS Estimators}

\begin{figure}[hb]
\centering
\begin{minipage}{0.45\textwidth}

 \includegraphics[width=0.99\textwidth]{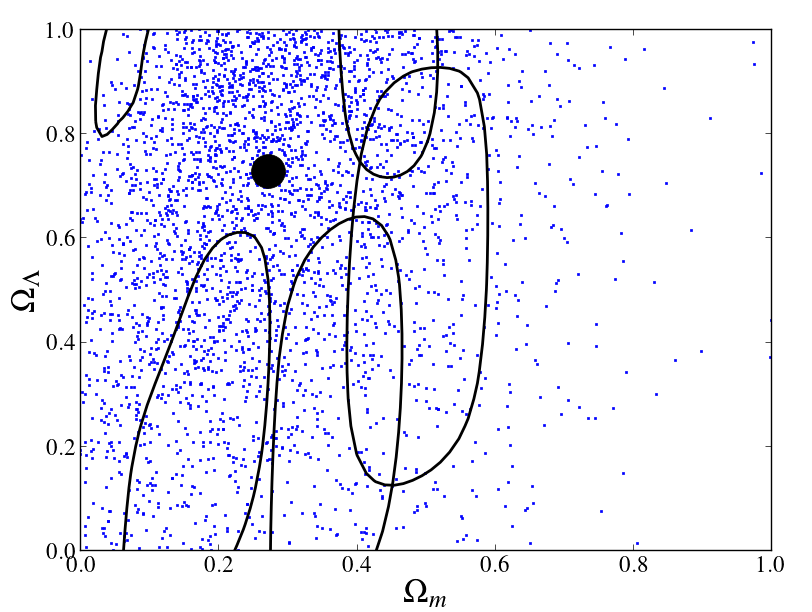}
\end{minipage}
\begin{minipage}{0.45\textwidth}

 \includegraphics[width=0.99\textwidth]{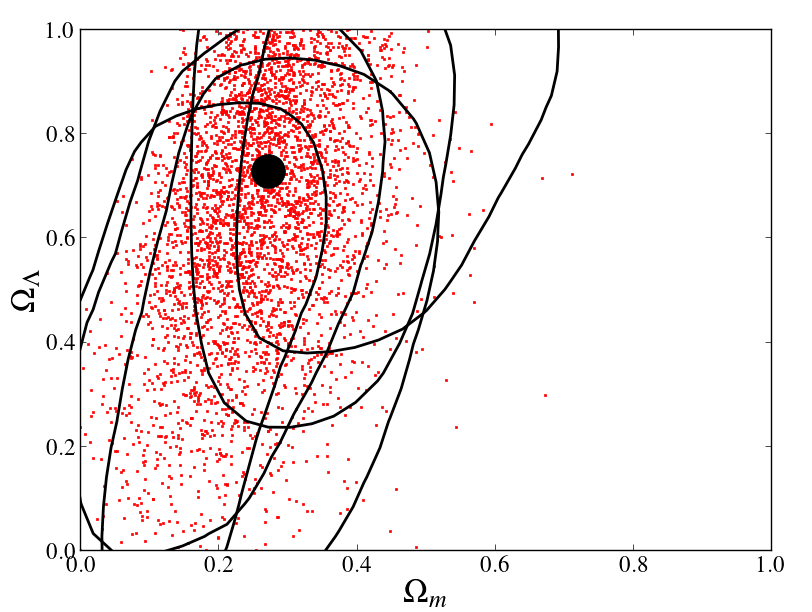} 
\end{minipage}
\caption{Scatter plots of the maximum posterior estimates for the cosmological parameters for each of 5000 correlated datasets using the wedding cake covariance matrix (eq. \ref{wedding}). \emph{Left panel:} BEAMS assuming the data are uncorrelated. \emph{Right panel:} correlated BEAMS with numerical marginalisation over covariance matrix parameters and types. Also plotted are the 95\% credible contours from five randomly selected datasets to give an idea of the size and shape of the contours. This shows that, while both methods are unbiased on average, correlated BEAMS is a much better estimator and has far superior coverage properties (88.2\% vs. only 7.2\%).}
\label{sim}

\end{figure}

To test the accuracy of the correlated BEAMS contours, we created 5000 datasets, each consisting of 200 correlated supernovae, using the wedding cake covariance matrix. We then ran a 100 000 step MCMC chain for both uncorrelated and correlated BEAMS on each of the datasets, totaling about a billion MCMC steps. Figure \ref{sim} shows scatter plots for both uncorrelated BEAMS and correlated BEAMS. Each point represents the maximum posterior values of the parameters for one of the 5000 datasets. We computed the mean squared error (MSE), which is the sample average of the squared distance between the estimates and the true value for the parameters (in this case $\Omega_m$ and $\Omega_{\Lambda}$), for both uncorrelated and correlated BEAMS. We found the relative efficiency, defined to be the ratio of the MSE for uncorrelated BEAMS to that of correlated BEAMS, to be 2.3, which implies that correlated BEAMS is a much more efficient estimator than uncorrelated BEAMS for correlated data, i.e. it estimates parameters with much less scatter.
\\
We also plotted the 95\% 
credible contours for five randomly selected datasets to show how the size and shape of the contours are underestimated by uncorrelated BEAMS but well estimated by correlated BEAMS, for these correlated datasets. To quantify this point, we computed the coverage, which we here define to be the proportion of the 5000 datasets where the true value (marked by the black circle in figure \ref{sim}) lies within the 95\% 
credible interval for each dataset derived from the corresponding MCMC chain for that dataset. We found that the coverage was 88.2\% for correlated BEAMS and only 7.2\% for uncorrelated BEAMS, showing that accounting for the correlations is crucial in getting the correct contours. Ideally, the coverage should be close to 95\%. However, the reader is cautioned that coverage is a frequentist concept and only for asymptotically large datasets would we expect the frequentist coverage to coincide with the corresponding coverage from Bayesian credible intervals. As our datasets only have 200 points each, we should not be surprised if the coverage is not exactly 95\%. Undercoverage occurs even for the standard $\chi^2$ method applied to supernovae datasets (see figure 8 in \cite{march}). 

\begin{figure}[hb]
\centering
\includegraphics[width=0.8\columnwidth]{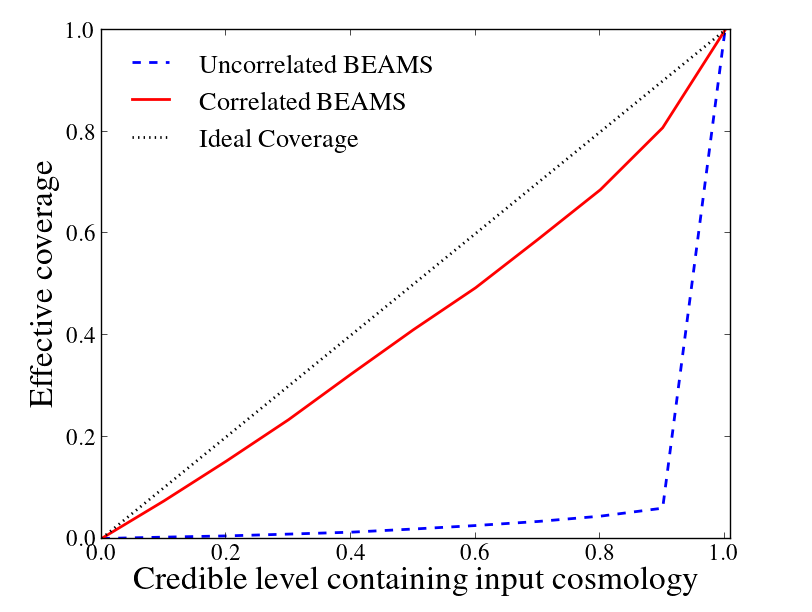}
\caption{Effective coverage, defined to be the fraction of datasets which contain the input cosmology at a given credible level for the 5000 datasets shown in figure \ref{sim} for both uncorrelated BEAMS (blue, dashed line), correlated BEAMS (red, solid line) and for the ideal coverage case (black, dotted line). While correlated BEAMS is close to ideal, implying that its contours can be trusted, uncorrelated BEAMS dramatically undercovers and hence the contours cannot be trusted for correlated data, as expected.}
\label{bias}
\end{figure}

The coverage properties of both estimators are illustrated in figure \ref{bias}. For both uncorrelated and correlated BEAMS we plotted the effective coverage (on the y-axis), i.e. the fraction of datasets for which the given credible contour contains the input cosmological parameters, at that level of credibility (on the x-axis). So for example, the 0.95 credible level corresponds to a coverage of 0.882 and 0.072 for correlated and uncorrelated BEAMS respectively. Ideal coverage is the diagonal straight line across the plot, where (for example) 0.95 of the datasets would contain the input cosmology in the 0.95 credible contour. It is clear that the coverage of correlated BEAMS is close to ideal and far superior
to that of uncorrelated BEAMS.

\section{A perturbative expansion of the BEAMS posterior}
\label{p_beams}
\vspace{-5pt}
\subsection{Theory}
For future surveys, in exactly the limit where it will be important to apply BEAMS to avoid biases from probability cuts, the light curve data for both candidates and templates will be excellent and the probability that most candidates are a Ia will therefore be close to zero or unity. This is illustrated in figure \ref{des}, which shows simulated DES probabilities using MLCS2k2 \cite{mlcs} and SNANA \cite{snana}. 
\\
\noindent
In this limit of abundant, high-quality data, we may expect we could perform a perturbative expansion of the full correlated BEAMS posterior, to find an approximation to the full posterior as an alternative to the numerical marginalisation over type. If we write eq. (\ref{beams_eq}) as a function of the probabilities, $\vec{p}$, we can Taylor expand the posterior around the point where the probabilities are rounded to either zero or one ($P_{Rnd}$): 
\begin{eqnarray}
P(\theta|D,\vec{p}) \propto P(\theta|D,\vec p)  \Big | _{P_{Rnd}} + \, \vec {\epsilon} \cdot \nabla P(\theta|D,\vec p)\Big | _{P_{Rnd}} + \, \frac{1}{2} \vec \epsilon \cdot \Big[ \vec\epsilon \cdot H(P(\theta|D,\vec p))\Big]\Big | _{P_{Rnd}} + \, \cdots, \qquad
\end{eqnarray}
\noindent
where $\epsilon_i=p_i - Rnd(p_i)$ and $Rnd(p_i)$ is the rounded value of the probability of the i'th data point (which will be either 0 or 1) and $H$ is the Hessian matrix. 
\\
Appendix \ref{appen} contains details of the derivation. The resulting posterior, up to the second order term, is
\begin{eqnarray}
P(\theta|D,\vec{\epsilon}) \propto P(\theta|D,\vec p)  \Big | _{P_{Rnd}}  &+& \, \displaystyle\sum\limits_{i=1}^{N} \bigg( \epsilon_i \times \left[P(D|\theta,\vec \tau_{i=\rm{Ia}}) - P(D|\theta,\vec \tau_{i=\rm{nIa}}) \right] \times P(\theta) \bigg)\, +
\nonumber
\\ &+&\frac{1}{2}\,\displaystyle \sum \limits_{i=1}^{N} \displaystyle \sum \limits_{\substack {j=1 \\ j \neq i}}^{N} \bigg(\epsilon_i \times \epsilon_j \times \Big[P(D|\theta,\vec \tau_{i,j=\rm{Ia}}) + P(D|\theta,\vec \tau_{i,j=\rm{nIa}}) \, - 
\nonumber
\\ 
   &-& P(D|\theta,\vec \tau_{i=\rm{Ia},j=\rm{nIa}}) - P(D|\theta,\vec \tau_{i=\rm{nIa},j=\rm{Ia}}) \Big] \Big[P(\theta) \Big]   \bigg) \,+\, \cdots. \qquad
\end{eqnarray}

\subsection{Results}
\begin{figure}[ht]
\vspace{-10pt}
  \centering
  \includegraphics[width=0.8\columnwidth]{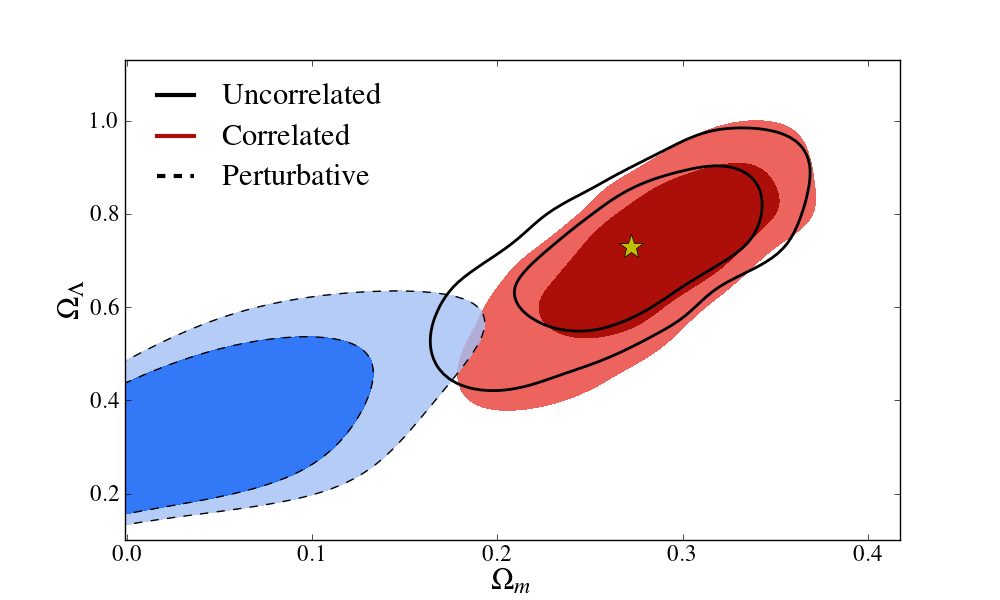}
  \caption{Comparison between uncorrelated BEAMS (line contours), correlated BEAMS (filled, red contours) and the perturbative expansion of BEAMS (filled blue contours with dashed lines), for an uncorrelated dataset, showing how the perturbative expansion fails badly even in the uncorrelated case.}
  \label{p-beams_uncor}
\end{figure}

\begin{figure}[ht]
\vspace{-10pt}
  \centering
  \includegraphics[width=0.8\columnwidth]{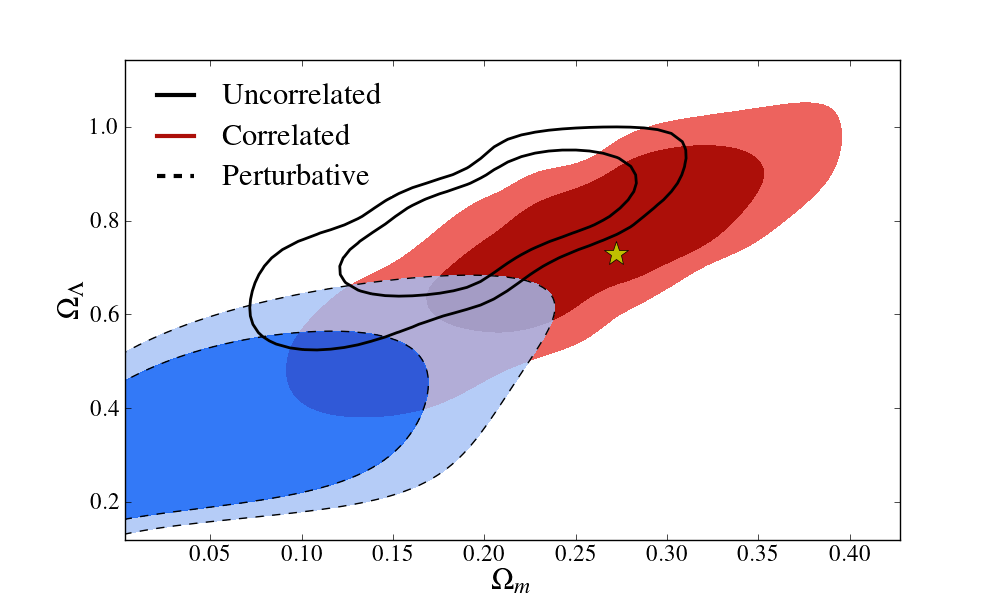}
  \caption{Comparison between uncorrelated BEAMS (line contours), correlated BEAMS (filled, red contours) and the perturbative expansion of BEAMS (filled blue contours with dashed lines), for a correlated dataset, using the decaying diagonal covariance matrix, showing how the perturbative expansion fails badly }
  \label{p-beams_cor}
\end{figure}

This perturbative approximation breaks down because we expand the posterior about the rounded off probabilities (that is, if the probability is close to $1$ we take the object to be a Ia and a nIa if it close to $0$), but due to the extreme nature of the likelihood, this tends to lie quite far from the maximum likelihood,
the point about which one would like to perform the expansion. This happens because $\sigma_{\text{Ia}}/\sigma_{\text{nIa}}$ is very small, hence if too many nIa's are mistyped as Ia's when rounding off the probabilities, these terms cause higher order terms in the Taylor expansion to dominate, however we only include terms up to second order term in this expansion in order for the analysis remain computationally viable (the second order term is an $N^3$ calculation, the third order $N^4$ and so on). Thus the perturbative expansion is not a good enough approximation of the true likelihood.
\\
 This can be seen in figures \ref{p-beams_uncor} and \ref{p-beams_cor}, where we show 1 and 2-$\sigma$ contours in the $\Omega_m - \Omega_\Lambda$ plane. In each case, the fiducial value is shown by a star. While the underlying data are correlated, the uncorrelated form of BEAMS \emph{assumes} uncorrelated data. In addition, perturbative BEAMS cannot sufficiently correct for the mistyped terms, and thus fail to recover the simulated data model at $> 3\sigma$. 

\begin{figure}[h!]

\begin{minipage}{0.5\textwidth}

 \includegraphics[width=0.99\textwidth]{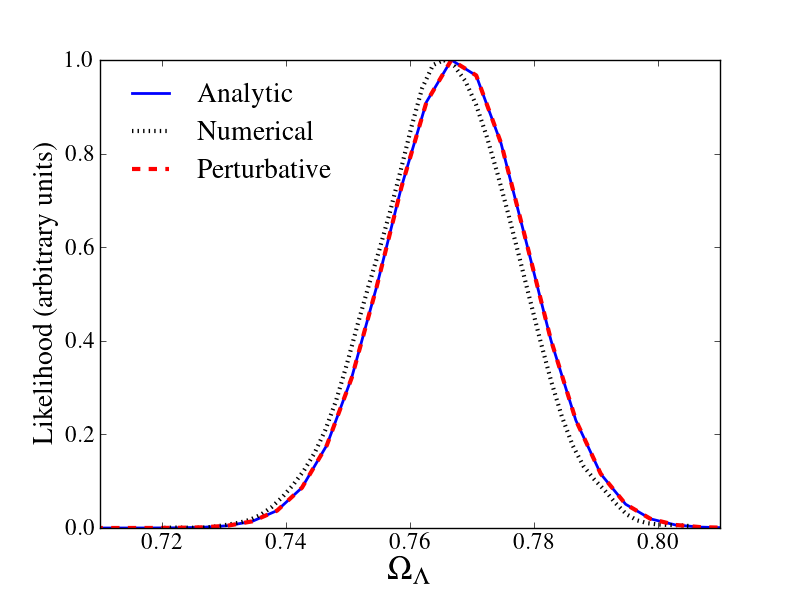}
\end{minipage}
\begin{minipage}{0.5\textwidth}

 \includegraphics[width=0.99\textwidth]{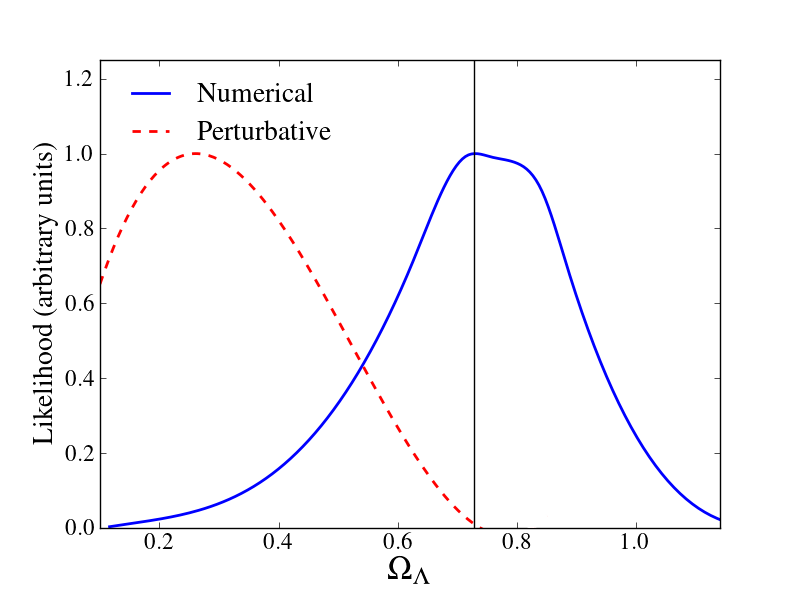}
\end{minipage}
\caption{These plots demonstrate the difference between the true BEAMS likelihood (dotted, black line), the perturbative approximation (dashed, red line) and correlated BEAMS with numerical marginalisation (solid, blue line). \emph{Left panel:} A set of ten data points was created using the decaying covariance matrix to correlate the data. $\Omega_m$ and $H_0$ are set to $0.272$ and $70.4$ respectively and $\Omega_\Lambda$ is allowed to vary. For this small dataset, perturbative BEAMS and correlated BEAMS are both good approximations to the true BEAMS likelihood. \emph{Right panel:} A similar dataset except with 200 data points this time. Now, perturbative BEAMS poorly approximates the true likelihood and produces a bias in $\Omega_\Lambda$. BEAMS with numerical marginalisation, however, is unbiased. The true likelihood cannot be calculated with such a large dataset.}
\label{fail_p_beams_small}

\end{figure}

\noindent
The failure of perturbative BEAMS to correct for mistyped terms can be understood further by considering figure \ref{fail_p_beams_small}. In the left panel, we show perturbative BEAMS successfully approximates the true likelihood, as does correlated BEAMS using numerical marginalisation, for a small dataset. The right panel, however, shows this is not the case for a large dataset. The perturbative BEAMS estimate for $\Omega_\Lambda$ is biased, because mistyped objects in the dataset cause higher order terms in the perturbative approximation to dominate. As shown previously, however, correlated BEAMS (with numerical marginalisation of types) accurately recovers the input cosmology.

\clearpage

\section{Conclusion}
\label{conclude}
Photometric supernova surveys with unprecedented amounts of data will provide an exciting opportunity to learn about the structure and evolution of the universe. 
Due to the vast number of supernova candidates and their extended redshift distribution in large future surveys, it will be difficult, if not impossible, to spectroscopically follow-up all candidates as is normally required in cosmological analyses. BEAMS is a rigorous statistical method which avoids biases while using all supernova candidates, together with the probability that a candidate is a Type Ia supernova, derived from the multicolour lightcurves of the candidate. Until now, BEAMS has been applied assuming the supernovae are uncorrelated \cite{kunz,hlozek,newling}, an assumption which will be inappropriate for future surveys. Without this assumption, the analytical form of the BEAMS posterior is computationally unfeasible. If the uncorrelated form of BEAMS is applied to a dataset with correlated systematic uncertainties, the posterior for the cosmological parameters can be incorrectly estimated.
\\
To deal with this 'exponential catastrophe', we have explored two different approaches. The first marginalises over all the possible combinations of object type numerically instead of analytically, by including the types as discrete nuisance parameters in our MCMC chains, making it computationally efficient. We have shown, with three separate models of covariance matrices, that this algorithm successfully recovers the input cosmology in the correlated case without bias. In addition, we have shown with 5000 mock datasets that the correlated BEAMS credible contours are reliable estimates of the true error contours. This is something that cannot be easily reproduced without using the correlated BEAMS formalism.
\\
 The second approach we considered was a perturbative expansion of the BEAMS posterior which typically fails because, when too many objects are mistyped, the higher order terms of the expansion (which are neglected due to computational constraints) dominate over the lower order terms, even when the probabilities are very close to zero or one thus producing a biased posterior. However, with numerical marginalisation over types, correlations between supernovae do not appear to be an impediment to using BEAMS 
in analysing future photometric supernova surveys.

\acknowledgments
We thank the anonymous referee for comments on the paper. Michelle Knights thanks Roberto Trotta for useful discussions and Cathy, Dee and Geoff  Knights for comments on the draft. Michelle Knights acknowledges support from the NRF/SKA and from the University of Cape Town. BB acknowledges the NRF and the SKA for funding. MV was supported by National Research Foundation grant 78870. MK acknowledges funding by the Swiss NSF. MS is funded by the SKA Project. Part of the numerical calculations for this paper were performed on the Andromeda cluster of the University of Geneva.

\appendix

\section{Perturbative BEAMS}
\label{appen}
\subsection{Theory}
The aim is to derive a formula for the BEAMS posterior in the case where the probabilities are all close to zero and one. We start with the BEAMS posterior in the general case:

\begin{eqnarray}
 P(\theta|D) = \displaystyle\sum\limits_{\tau}P(D|\theta,\vec \tau)P(\theta)P(\vec \tau) / P(D)
\end{eqnarray}
\noindent
This results in $2^N$ terms being calculated, in the case where there are two types.
\\
We can Taylor expand this posterior, written as a function of the probabilities $\vec{p}$ (this is a length-$N$ vector containing the probability for each object), around the point where the probabilities are rounded to either zero or one. We define
the vector $\vec \epsilon$ such that $\epsilon_i=p_i - Rnd(p_i)$, where $Rnd(p_i)$ is the rounded value of the probability of the i'th data point (which will
be either 0 or 1). So for example, if $p_i=0.99$ then $\epsilon_i=-0.01$ and if $p_i=0.02$ then $\epsilon_i=0.02$ etc. 
\\
The BEAMS posterior becomes:

\begin{eqnarray}
P(\theta|D,\vec{p}) \propto P(\theta|D,\vec p)  \Big | _{P_{Rnd}} + \, \vec {\epsilon} \cdot \nabla P(\theta|D,\vec p)\Big | _{P_{Rnd}} + \, \frac{1}{2} \vec \epsilon \cdot \Big[ \vec\epsilon \cdot H(P(\theta|D,\vec p))\Big]\Big | _{P_{Rnd}} + \, \cdots, \qquad
\end{eqnarray}
\noindent
where H is the Hessian matrix. 

\subsubsection{The first order term}
Now we will explicitly calculate the first order derivative term. The i'th component of $\nabla P(\theta|D,\vec p)$ is given by 

\begin{eqnarray}
\frac {\partial P(\theta|D,\vec{p})}{\partial p_i}  = \displaystyle\sum\limits_{\tau} P(D|\theta,\vec \tau) P(\theta) \frac{\partial P(\vec \tau)}{\partial p_i}.
\end{eqnarray}
\noindent
This derivative is trivial since $P(\vec \tau)$ is linear in the $p_i$'s. $P(\vec \tau)$ is defined as

\begin{eqnarray}
\label{P}
P(\vec \tau) = \displaystyle\prod\limits_{\tau_j=\rm{Ia}} p_j \displaystyle\prod\limits_{\tau_k=\rm{nIa}}(1-p_k)\,.
\end{eqnarray}
\noindent
Since only one of these terms is a function of $p_i$ (where i equals either j or k depending on its type), the derivative is given by

\begin{eqnarray}
\frac{\partial P(\vec \tau)}{\partial p_i} = \frac{\partial P_{\tau_i}}{\partial p_i} P(\vec \tau_{-i}).
\end{eqnarray}

\noindent
Here, $P(\vec \tau_{-i})$ is equivalent to eq. (\ref{P}), with the contribution to the product from i'th object's probability removed. Since $P_{\tau_i}$ is either $p_i$
or $(1-p_i)$, its derivative is given by

\begin{equation}
\frac{\partial {P_{\tau_i}}}{\partial {p_i}} = 
\begin{cases}   
1 & \text{if $\tau_i=\rm{Ia}$} \\
-1 & \text{if $\tau_i=\rm{nIa}$}.
\end{cases}
\end{equation}
\noindent
Thus:
\begin{equation}
\frac{\partial {P(\theta|D,\vec{p})}}{\partial {p_i}} = \displaystyle\sum\limits_{\tau} \pm P(D|\theta,\vec \tau) P(\theta) P(\vec \tau_{-i})\,.
\end{equation}

\noindent
Once we subsitute for the rounded probabilities, many of the terms in $P(\vec {\tau}_{-i})$ will go to zero. In fact, each element of $\vec{\tau}_{-i}$ must equal its rounded type for $P(\vec {\tau}_{-i})$ in order to be non-zero, in which case, $P(\vec{\tau}_{-i}) = 1$. This leaves only two terms, corresponding to the two possible types of the i'th data point. Thus:

\begin{equation}
 \frac{\partial {P(\theta|D,\vec{p})}}{\partial {p_i}} \bigg | _{P_{Rnd}} = \left[P(D|\theta,\vec \tau_{i=\rm{Ia}}) - P(D|\theta,\vec \tau_{i=\rm{nIa}}) \right] P(\theta)\,,
\end{equation}
\noindent
where $\vec {\tau}_{i=\rm{Ia}}$ is the same as $\vec {\tau}_{Rnd}$ (where $\tau_{Rnd_i}=\rm{Ia}$ if $Rnd(p_i)=1$ and $\tau_{Rnd_i}=\rm{nIa}$ if $Rnd(p_i)=0$) with the i'th element being Type Ia
and $\vec \tau_{i=\rm{nIa}}$ is the same except the i'th element is Type nIa. 
\\
\noindent

Finally, the first order BEAMS posterior is given by
\begin{eqnarray}
P(\theta|D,\vec{\epsilon}) \propto P(\theta|D,\vec p)  \Big | _{P_{Rnd}}  + \, \displaystyle\sum\limits_{i=1}^{N} \bigg( \epsilon_i \times \left[P(D|\theta,\vec \tau_{i=\rm{Ia}}) - P(D|\theta,\vec \tau_{i=\rm{nIa}}) \right] \times P(\theta) \bigg)\,.
\end{eqnarray}
\noindent
For N data points, the order of this calculation is $2N$ times the order of the likelihood calculation. So in the case where the
likelihood is given by the usual Gaussian form $e^{-\chi^2/2}$, this would be an order $2N^2$ calculation.

\subsubsection{The second order term}
Our treatment can be repeated for the second order term. The second order derivative is given by
\begin{equation}
\label{post}
 \frac{\partial^2 {P(\theta|D,\vec{p})}}{\partial {p_i}\partial {p_j}} = \displaystyle\sum\limits_{\tau} P(D|\theta,\vec \tau) P(\theta) \frac{\partial^2 P(\vec \tau)}{\partial p_i \partial p_j}\,.
\end{equation}

\noindent 
$P(\vec{\tau})$ can again be expressed as a product of one $i$-dependent term, one $j$-dependent term and one term independent of both
\begin{equation}
 P(\vec{\tau}) = P_{\tau_i} P_{\tau_j} P_{\tau_{-i,j}}\,,
\end{equation}
if $i \neq j$. If $i=j$, then the second derivative of $P(\vec{\tau})$ will be zero.
\\
By the product rule, this is

\begin{equation}
 \frac{\partial^2 {P(\vec \tau)}}{\partial {p_i}\partial {p_j}} = \frac{\partial_{\tau_i}}{\partial p_i} \frac{\partial_{\tau_j}}{\partial p_j} P(\vec \tau_{-i,j})\,.
\end{equation}
\noindent
Which evaluates to
\begin{equation}
\frac{\partial^2 {P(\vec \tau)}}{\partial {p_i}\partial {p_j}} = 
\begin{cases}   
0 & \text{if i=j} \\
1 & \text{if $i \neq j$ and $\tau_i=\tau_j$} \\
-1 & \text{if $i \neq j$ and $\tau_i \neq \tau_j$}.
\end{cases}
\end{equation}
\noindent
Similarly to the first order term, when we evaluate the full posterior (eq. (\ref{post})) at the rounded probabilities, we find that
only four terms remain (still for $i \neq j$):
\begin{eqnarray}
 \left( \frac{\partial^2 {P(\theta|D,\vec{p})}}{\partial {p_i}\partial {p_j}} \right) \Big|_{P_{Rnd}} = & & \Big[P(D|\theta,\vec \tau_{i,j=\rm{Ia}}) + P(D|\theta,\vec \tau_{i,j=\rm{nIa}}) \, - \nonumber \\
& & \, - P(D|\theta,\vec \tau_{i=\rm{Ia},j=\rm{nIa}}) - P(D|\theta,\vec \tau_{i=\rm{nIa},j=\rm{Ia}}) \Big] \Big[P(\theta) \Big],
\end{eqnarray}
\noindent
where
\begin{align*}
 \vec \tau_{i,j=\rm{Ia}} = \vec \tau_{Rnd} \quad &\text{except with} \quad \tau_i=\tau_j=\rm{Ia}\\
 \vec \tau_{i,j=\rm{nIa}} = \vec \tau_{Rnd} \quad &\text{except with} \quad \tau_i=\rm{Ia},\tau_j=\rm{nIa}\\
 \vec \tau_{i=\rm{Ia},j=\rm{nIa}} = \vec \tau_{Rnd} \quad &\text{except with} \quad \tau_i=\rm{nIa},\tau_j=\rm{Ia}\\
 \vec \tau_{i=\rm{nIa},j=\rm{Ia}} = \vec \tau_{Rnd} \quad &\text{except with} \quad \tau_i=\tau_j=\rm{nIa}.
\end{align*}

\noindent
Putting this all together, the second order term of the BEAMS posterior is
\begin{eqnarray}
 \frac{1}{2} \vec \epsilon \cdot \Big[ \vec\epsilon \cdot H(P(\theta|D,\vec p))\Big]\Big | _{P_{Rnd}} = & &\frac{1}{2}\,\displaystyle \sum \limits_{i=1}^{N} \displaystyle \sum \limits_{\substack {j=1 \\ j \neq i}}^{N} \bigg(\epsilon_i \times \epsilon_j \times \Big[P(D|\theta,\vec \tau_{i,j=\rm{Ia}}) + P(D|\theta,\vec \tau_{i,j=\rm{nIa}}) \, - \qquad \quad \nonumber \\ 
 & & \, - P(D|\theta,\vec \tau_{i=\rm{Ia},j=\rm{nIa}}) - P(D|\theta,\vec \tau_{i=\rm{nIa},j=\rm{Ia}}) \Big] \Big[P(\theta) \Big]   \bigg) .
\end{eqnarray}

\noindent
Thus the second order term is order $(4N)^2$ times the order of the likelihood calculation. For a Gaussian likelihood, this is an order $N^3$ calculation. 

\renewcommand{\refname}{\vspace*{-8mm}}{\section*{References}}
\bibliographystyle{JHEP}
\bibliography{beams_refs}

\end{document}